\documentclass[aps,numerical,superscriptaddress,floatfix,reprint,groupedaddres,longbibliography]{revtex4-2}
\usepackage{graphicx,color}
\usepackage{amsmath,amssymb,amsfonts,bm}
\usepackage[colorlinks=true,linkcolor=blue,plainpages=false]{hyperref}
\UseRawInputEncoding 

\newcommand{\be}{\begin{equation}}
\newcommand{\ee}{\end{equation}}
\newcommand{\ba}{\begin{eqnarray}}
\newcommand{\ea}{\end{eqnarray}}
\newcommand{\ban}{\begin{eqnarray*}}
\newcommand{\ean}{\end{eqnarray*}}

\def\v2{\mbox{$v_2$}}

\begin{document}

\title{Relativistic Viscous Hydrodynamics with Angular Momentum}
\medskip

\author{Duan She}
\affiliation{ Key Laboratory of Quark and Lepton Physics(MOE) and Institute of Particle Physics, central China Normal University , Wuhan, 430079, China.}
\affiliation{ Physics Department and Center for Exploration of Energy and Matter,
Indiana University, 2401 N Milo B. Sampson Lane, Bloomington, IN 47408, USA.}
\affiliation{Department of Modern Physics, University of Science and Technology of China, Anhui 230026, China}
\author{Anping Huang}
\affiliation{ Physics Department and Center for Exploration of Energy and Matter,
Indiana University, 2401 N Milo B. Sampson Lane, Bloomington, IN 47408, USA.}
\affiliation{School of Nuclear Science and Technology, University of Chinese Academy of Sciences, Beijing 100049, China.}

 \author{Defu Hou}
 \thanks{Corresponding author}
 \email{houdf@mail.ccnu.edu.cn}
  \affiliation{ Key Laboratory of Quark and Lepton Physics(MOE) and Institute of Particle Physics, central China Normal University , Wuhan, 430079, China.}

 \author{Jinfeng Liao}
 \thanks{Corresponding author}
\email{liaoji@indiana.edu}
\affiliation{ Physics Department and Center for Exploration of Energy and Matter,
Indiana University, 2401 N Milo B. Sampson Lane, Bloomington, IN 47408, USA.}

 \date{\today}


\begin{abstract}

\end{abstract}

\maketitle


Hydrodynamics   is a general theoretical framework for describing the long-time large-distance behaviors of  macroscopic physical systems. It has many important applications in various branches of physics, from cosmic expansion and galaxy/star evolutions at the very large scales to relativistic nuclear collisions at the very small scales. The core of hydrodynamics is about physical quantities protected by exact conservation laws, such as energy, momentum and conserved charges.  Past hydrodynamic studies almost entirely focus on the energy-momentum conservation and charge conservation. Only very recently, there has been a rapidly  increasing interest in understanding the role of angular momentum conservation in the hydrodynamic context and its implications for spin transport of underlying constituents. Such interest is strongly fueled by experimental observations of spin polarization   in rotating matter, with examples ranging from  condensed matter flow systems to subatomic fluids in relativistic nuclear collisions~\cite{STAR:2017ckg,ALICE:2019aid,STAR:2022fan,HADES:2022enx,2016NatPh..12...52T,2020NatCo..11.3009T}.

Active efforts are underway to develop a hydrodynamic theory framework for describing such systems, see e.g. ~\cite{Becattini:2009wh,Florkowski:2017ruc,Hattori:2019lfp,Fukushima:2020ucl,Bhadury:2020cop,Shi:2020htn,Li:2020eon,Weickgenannt:2020aaf,Fu:2021pok,Becattini:2021iol,Gallegos:2021bzp}.
Important progress has been made along this direction, while there also appear both conceptual and technical challenges especially in the  relativistic regime where the separation between spin and orbital components becomes subtle and confusions arise about  the property of energy-momentum tensor as well as the hydrodynamic gradient expansion.

\begin{figure}[!hbt]
	\begin{center}
		\includegraphics[width=2.7in]{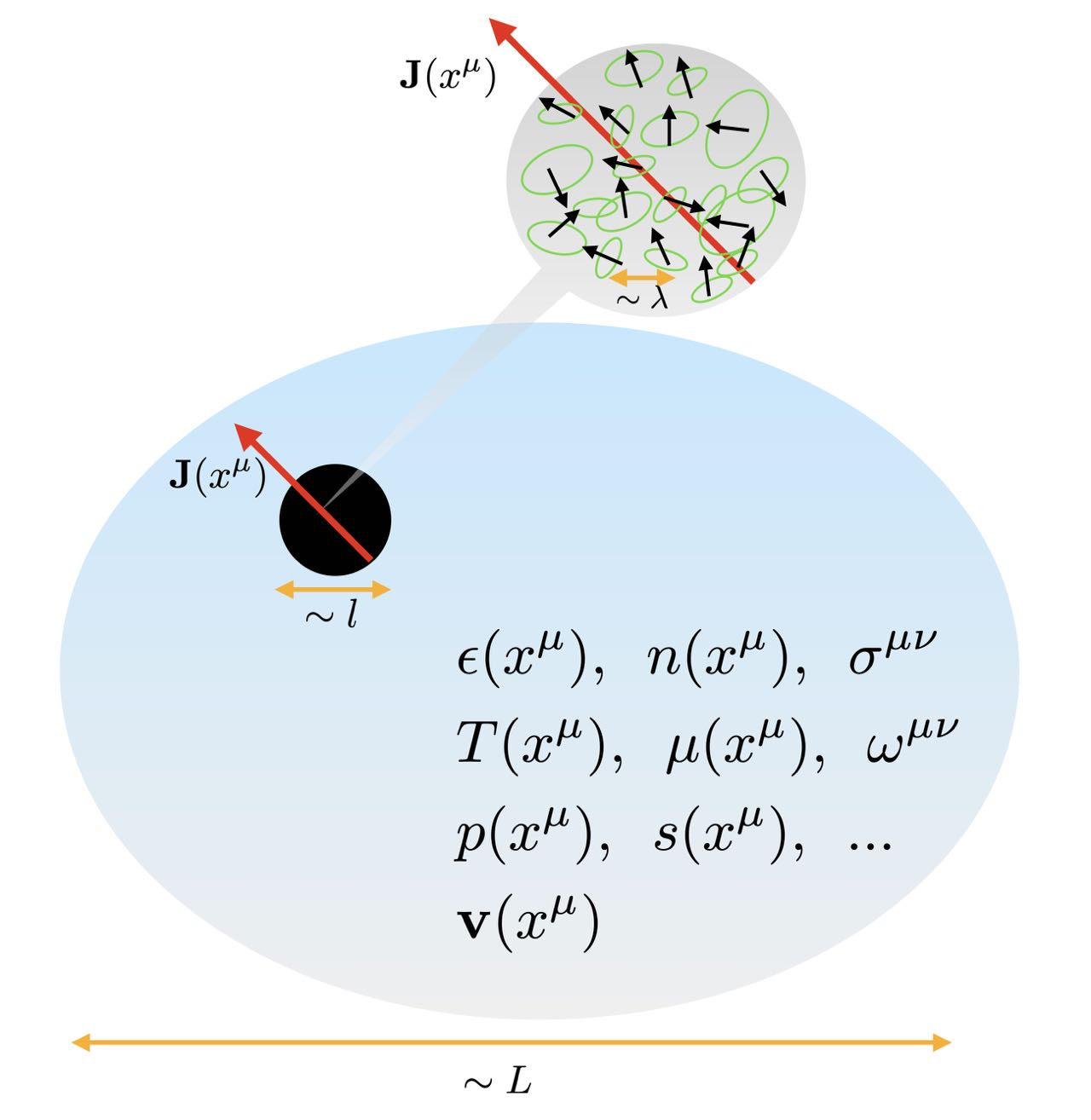}
		\caption{An illustration of the hydrodynamic regime. The bulk behavior of the system, on the macroscopic scale $L$, is described by a set of hydrodynamic fields/variables such as temperature $T(x^\mu)$, charge density $n(x^\mu)$, fluid velocity $\mathbf{v}(x^\mu)$, etc. Individual fluid cells, on the hydrodynamic scale $l $, are considered to be  close to local thermal equilibrium. }
		\vspace{-0.5cm}
		\label{fig_1}
	\end{center}
\end{figure}

Let us begin with a conceptual discussion on the hydrodynamic description of a general fluid system. One starts by assuming a separation between the macroscopic scale $L$ (e.g. the system size) and the microscopic scale $\lambda$, which is determined by underlying dynamical interactions relevant for the thermal relaxation and equilibration among both spin and orbital angular momentum of the fluid constituents. This allows introducing an intermediate hydrodynamic scale $l$ for defining local fluid cells, with $\lambda \ll l \ll L$, a coarse-graining process as illustrated in Fig.~\ref{fig_1}. Each fluid cell is supposed to  be  close to local thermal equilibrium and can be represented by locally-defined hydrodynamic fields/variables such as temperature $T(x^\mu)$ (or equivalently energy density $\epsilon(x^\mu)$), chemical potential $\mu(x^\mu)$ (or charge density $n(x^\mu)$), pressure $p(x^\mu)$, entropy density $s(x^\mu)$, fluid velocity $\mathbf{v}(x^\mu)$, etc.

Hydrodynamic equations are nothing but macroscopic conservation laws for such hydrodynamic fields. For example, the usual relativistic hydrodynamics consists of the following equations for energy-momentum tensor $T^{\mu\nu}$ as well as charge current $N^{\mu}$:
\begin{eqnarray}
&&\partial_{\mu}T^{\mu\nu}=0, \label{eq01}\\
&&\partial_{\mu}N^{\mu}=0, \label{eq02}
\end{eqnarray}
which are essentially continuity equations for transport currents pertaining to energy, momentum and charge. The remaining task then is to obtain constitutive relations, i.e. to express $T^{\mu\nu}$ and  $N^{\mu}$ in terms of various hydrodynamic fields (e.g. $\mathbf{v}, T, n,...$) and their derivatives in a systematic order-by-order gradient expansion.



Along the same line of consideration, let us examine the inclusion of angular momentum  by adding the corresponding  conservation equation:
\begin{eqnarray}
 \partial_{\mu}J^{\mu\alpha\beta}=0, \label{eq03}
\end{eqnarray}
where the tensor $J^{\mu\alpha\beta}$ is the local current associated with angular momentum transport. It should contain all contributions such as spin and orbital angular momentum.

To express the angular momentum carried by a local fluid cell is however tricky, with two contributions:
\begin{eqnarray}
J^{\mu\alpha\beta}=\left(x^{\alpha}T^{\mu\beta}-x^{\beta}T^{\mu\alpha}\right)+\Sigma^{\mu\alpha\beta} , \label{eq04}
\end{eqnarray}
where both $J^{\mu\alpha\beta}$ and $\Sigma^{\mu\alpha\beta}$ are anti-symmetric in $\alpha \leftrightarrow \beta$.
The first part, in the parentheses, comes from the angular momentum associated with the orbital motion of the fluid cell as {\em a ``whole object''}. Note the $T^{\mu\nu}$ here is the canonical energy-momentum tensor which could in principle have both symmetric and antisymmetric components: $T^{\mu\nu}\equiv T_{(s)}^{\mu\nu}+T_{(a)}^{\mu\nu}$.
 The second part, i.e. $\Sigma^{\mu\alpha\beta}$, counts all the {\em internal angular momentum} from within the fluid cell. Microscopically it includes both orbital and spin contributions from all microscopic constituents in that cell, as illustrated in Fig.~\ref{fig_1}.


The conserved quantity related to $\Sigma^{\mu\alpha\beta}$ is introduced as the antisymmetric tensor field $\sigma^{\alpha \beta}(x^\mu)$ which should represent the local angular momentum density, akin to the local charge density.   It should also have a corresponding ``Lagrangian multiplier'', the angular momentum chemical potential $\omega_{\alpha \beta}(x^\mu)$.
It shall be noted that the $\sigma^{\alpha \beta}$ and $\omega_{\alpha \beta}(x^\mu)$ should not be viewed as just spin tensor or spin chemical potential.

As is well known, hydrodynamic equations need to be ``closed'' with thermodynamic constraints that the hydrodynamic variables must satisfy.
Including angular momentum in the same way as other conserved quantities, one arrives at the following generalized conditions according to  the first law of thermodynamics:
 \begin{eqnarray} \label{eq_thermal}
&&  \epsilon=-p+Ts+\mu n+\omega_{\alpha\beta} \sigma^{\alpha\beta} \,  .
\end{eqnarray}
Furthermore, the second law of thermodynamics requires that the entropy can not decrease, i.e.
\begin{eqnarray}
\partial_{\mu}S^{\mu}\geq0, \label{eq05}
\end{eqnarray}
where $S^{\mu}=p\beta^{\mu}+\beta_{\nu}T^{\mu\nu}-\alpha N^{\mu}-\beta\omega_{\alpha\beta}\Sigma^{\mu\alpha\beta}$ is the entropy current. Additionally, one can use the thermodynamic equation of state to express e.g. $\epsilon, p, s, ...$ in terms of $T,\mu,\omega_{\alpha\beta}$.

The next key step is to find the constitutive relations that specify conserved quantities, i.e. $T^{\mu \nu}$, $N^\mu$ and $\Sigma^{\mu\alpha\beta}$ in terms of hydrodynamic variables. This can be done through a systematic gradient expansion. One starts by assuming perfect local thermal equilibrium to write down them entirely in terms of local variables and obtain the ideal hydrodynamics. One next introduces viscous terms into constitutive relations, involving only single-derivative terms of hydrodynamic variables to obtain Navier-Stokes type of viscous hydrodynamics. The procedure can be further carried on by including higher-order-derivative terms.

In the ideal hydrodynamic limit, all conserved quantities can be uniquely determined from local variables.
The energy-momentum tensor and charge current are given by the familiar forms:
$T_{\left(0\right)}^{\mu\nu}= \epsilon u^{\mu}u^{\nu}-p\Delta^{\mu\nu},\,
N_{\left(0\right)}^{\mu}=nu^{\mu}  $,
where $u^\mu = (\gamma, \gamma \mathbf{v})$ (with $\gamma=1/\sqrt{1+\mathbf{v}^2}$).
For angular momentum, one  similarly obtains
$\Sigma_{(0)}^{\mu\alpha\beta}=\sigma^{\alpha\beta}u^{\mu}$.
Combining this with Eqs.~(\ref{eq01},\ref{eq02},\ref{eq03}), one thus obtains ideal hydrodynamic equation for the angular momentum~\cite{Becattini:2009wh,Florkowski:2017ruc}:
\begin{eqnarray}
\partial_{\mu}J_{(0)}^{\mu\alpha\beta}=\sigma^{\alpha\beta}\theta + D \sigma^{\alpha\beta} = 0,
\label{eq08}
\end{eqnarray}
where $\theta=\partial_\mu u^\mu$ and  $D=u_{\nu}\partial^{\nu}$. At this order,  the constraint (\ref{eq05}) is saturated, i.e.   $\partial_{\mu}S_{\left(0\right)}^{\mu}=0$ as it should be: no entropy is generated in ideal hydrodynamics.

The first  dissipative corrections can be added by including viscous currents constructed from linear order of gradient terms, a procedure known to become frame-dependent.
In the so-called Eckart frame defined by charge current $N^{\mu}$, one can carry out a general and systematic decomposition  of the physical currents $N^\mu$, $T^{\mu\nu}$ and $\Sigma^{\mu\alpha\beta}$ into longitudinal and transverse components:
\begin{eqnarray}
&&T^{\mu\nu}= \epsilon u^{\mu}u^{\nu}-\left(p+\Pi\right)\Delta^{\mu\nu}+2u^{(\mu}q^{\nu)}+\pi^{\mu\nu},\label{eq19}\\
&&N^{\mu}=nu^{\mu},\label{eq20}\\
&&\Sigma^{\mu\alpha\beta}=u^{\mu} \sigma^{\alpha\beta}  +2u^{[\alpha}\Delta^{\mu\beta]}\Phi
\nonumber \\
&& \qquad \qquad \qquad \quad
+ 2u^{[\alpha}\tau_{(s)}^{\mu\beta]}+ 2u^{[\alpha}\tau_{(a)}^{\mu\beta]}+\Theta^{\mu\alpha\beta}.
\label{eq21}
\end{eqnarray}
In the above, we recognize the familiar bulk viscous pressure $\Pi$,  shear viscous tensor $\pi^{\mu\nu}$ and the diffusion flux  $q^\mu$. Furthermore,  there emerge new dissipative quantities $\Phi,  \tau_s^{\mu\beta}, \tau_a^{\mu\beta}, \Theta^{\mu\alpha\beta}$ associated with dissipative processes of angular momentum transport in Eq.(\ref{eq21}).
 These viscous terms are all considered linear-order terms in gradient expansion.
  Note for a pair of Lorentz indices, we use notation $^{(\mu ... \nu)}$ to indicate symmetrizing  while $^{[\mu ... \nu]}$ to indicate anti-symmetrizing.  Later we will also use notation $^{\langle \mu ... \nu \rangle}$  for the symmetric traceless projection.


Finally one considers the constraint Eq.~(\ref{eq05}) for the entropy current  $ S^{\mu}$, the divergence of which must be non-negative  for all possible fluid configurations (aka ``entropy never decreases''). The only way for this to happen is for all contributing terms   to be space-like quadratic terms, which helps uniquely fixing various viscous terms to leading order. For energy-momentum and charge current, the results are similar to conventional Navier-Stokes hydrodynamics:
\begin{eqnarray}
&&\Pi=-\zeta\theta,\label{eq22}\\
&&\pi^{\mu\nu}=2\eta\nabla^{\langle\mu}u^{\nu\rangle},\label{eq23}\\
&&q^{\mu} =\kappa T\left(\frac{\nabla^{\mu}T}{T}-Du^{\mu}\right) \nonumber \\
&& \quad \,\,
 =-\frac{\kappa nT^{2}}{\epsilon+p}\left[\nabla^{\mu}\left(\frac{\mu}{T}\right)+\frac{\sigma^{\alpha\beta}}{n}\nabla^{\mu}\left(\frac{\omega_{\alpha\beta}}{T}\right)\right],\label{eq24}
 \end{eqnarray}
 where the positive coefficients $\zeta, \eta, \kappa$ are the familiar  bulk viscosity, shear viscosity and heat conductivity, respectively. For the angular momentum part, new results are obtained for its various viscous components:

 \begin{eqnarray}
&&\Phi=-\chi_1  u^{\alpha}\nabla^{\beta}\left(\frac{\omega_{\alpha\beta}}{T}\right),\label{eq25}\\
&&\tau_{(s)}^{\mu\beta}=-\chi_2  u^{\alpha} \bigg{[} \left(\Delta^{\beta\rho}\Delta^{\mu\gamma}+\Delta^{\mu\rho}\Delta^{\beta\gamma}\right) \nonumber \\
&& \qquad \qquad \qquad \qquad -\frac{2}{3}\Delta^{\mu\beta} \Delta^{\rho\gamma}\bigg{ ] }
\nabla_{\gamma}\left(\frac{\omega_{\alpha\rho}}{T}\right), \label{eq26} \\
&&
\tau_{(a)}^{\mu\beta}=- \chi_3 u^{\alpha}\left(\Delta^{\beta\rho}\Delta^{\mu\gamma}-\Delta^{\mu\rho}\Delta^{\beta\gamma}\right)\nabla_{\gamma}\left(\frac{\omega_{\alpha\rho}}{T}\right),\label{eq27}\\
&&\Theta^{\mu\alpha\beta}= - \chi_{4} \left(u^{\beta}u^{\rho}\Delta^{\alpha\delta}-u^{\alpha}u^{\rho}\Delta^{\beta\delta}\right)\Delta^{\mu\gamma}\nabla_{\gamma}\left(\frac{\omega_{\delta\rho}}{T}\right) \nonumber \\
&& \qquad \qquad   +\chi_{5} \Delta^{\alpha\delta}\Delta^{\beta\rho}\Delta^{\mu\gamma}\nabla_{\gamma}\left(\frac{\omega_{\delta\rho}}{T}\right).
\label{eq28}
\end{eqnarray}
Here, five new positive transport coefficients$\chi_{1}, \chi_{2}, \chi_{3}, \chi_{4}$ and $\chi_{5}$ are identified, which quantify the angular momentum diffusion in various modes.

One may notice a nontrivial contribution to the viscous heat current from the angular momentum gradient term in (\ref{eq24}).  It is also evident from (\ref{eq25}--\ref{eq27}) that temperature gradient would contribute to the viscous current of angular momentum transport. An interesting quantity much discussed in literature is the polarization pseudo-vector, which may be connected with the angular momentum current via $\Pi^\mu \sim \epsilon^{\mu\nu\rho\sigma} \Sigma_{0\nu\rho} T_{0\sigma}$, which would receive viscous contributions from angular momentum and temperature gradients as well as from bulk and shear gradient terms in the stress tensor. Note that if   the Landau frame or energy frame is used instead of the Eckart frame,  one obtains formally  the same expressions for the set of viscous constitutive relations in Eqs.(\ref{eq22}--\ref{eq27}) albeit with all quantities defined in the Landau frame.

As of this point, a framework of relativistic viscous hydrodynamics with angular momentum is developed as a most natural extension  of the original Navier-Stokes analysis into a new regime. In particular, the viscous terms for angular momentum transport have been found with five pertinent transport coefficients newly introduced. There remain however a number of highly intriguing conceptual issues  that have been  investigated in recent literature and still await a complete understanding.


One subtle question is whether the spin and orbital components of a fluid cell could be meaningfully separated thus allowing the spin portion to be treated as an independent hydrodynamic field. There is an obvious desire to do so, given that nuclear collision experiments measure the spin polarization of produced particles. Many studies~\cite{Hattori:2019lfp,Fukushima:2020ucl,Bhadury:2020cop,Li:2020eon,Gallegos:2021bzp} attempt to build the ``spin hydrodynamics'' based on such a scenario. The separation may not be unique with different choices related to pseudo-guage transformation~\cite{Fukushima:2020ucl}. Further concerns arise in the context of hydrodynamic coarse-graining, and there is no guarantee that the scale of spin tensor variation could be cleanly separated from microscopic scale as well as from the scale of fluid orbital motion.
In fact, if spins from different cells could indeed correlate in a hydrodynamic way, it is difficult to imagine that these spins would not interact with local orbital motion. Furthermore, one may note that a general hydrodynamics with angular momentum shall encompass not only quantum field systems with spins but also those without spins but possessing nonzero angular momentum.

Another tricky question is about the power counting of newly introduced angular momentum variables like $\omega_{\alpha\beta}$ and $\sigma^{\alpha\beta}$. In terms of gradient expansion, the framework obtained above consistently counts both
$\sigma^{\alpha \beta}$ and $\omega_{\alpha \beta}$ as zeroth order terms. This might not be the only possible choice. To make an analogy, in the context of magnetohydrodynamics, different formulations arise from different counting of the magnetic field terms.  For the angular momentum,  a number of works (e.g. \cite{Hattori:2019lfp,Fukushima:2020ucl,Li:2020eon}) treat the $\omega_{\alpha \beta}$ as first-order term in gradient expansion, resulting in different forms of the viscous contributions. On the other hand, this may cause concern about consistency with   thermodynamic relation Eq.~(\ref{eq_thermal}), which are meant to relate locally defined quantities for fluid cells in local thermal equilibrium. Such quantities   are considered as zeroth order terms in the hydrodynamic context and the physical meaning is unclear if one were to mix terms of different orders in this relation.

There are many interesting problems in the further development of relativistic viscous hydrodynamics with angular momentum. One obvious example is the construction of second-order casual framework beyond the Navier-Stokes analysis above. See recent developments in e.g. \cite{Li:2020eon,Gallegos:2021bzp,Weickgenannt:2022zxs}, using either macroscopic analysis   or microscopic approach. The microscopic approach  starts from transport equations with collisions terms for instance in \cite{Wang:2020pej, Lin:2021mvw}. From a microscopic point of view, the dynamics underlying the conversion between orbital and spin angular momentum can be understood in terms of particle collisions. The nonlocal collision term is of second order in $\hbar$, which is responsible for the conversion of orbital to spin angular momentum\cite{Weickgenannt:2020aaf,Wang:2020pej, Lin:2021mvw} . Computation of the  newly found transport coefficients here for angular momentum transport would be desirable and can be done by starting with e.g. quantum transport theories~\cite{Bhadury:2020cop,Shi:2020htn, Weickgenannt:2020aaf}. Concerning frame choice, the presence of the angular momentum current brings the possibility of a new choice for local rest frame and its implication has yet to be understood. Given the new hydrodynamic variables brought by the angular momentum, various new hydrodynamic modes akin to usual sound waves are expected and can be revealed by a linearization analysis of the hydrodynamic equations involving angular momentum variables. A fluid system made of massless fermions deserves special interest, in which anomalous transport such as chiral magnetic and vortical effects would occur. Such a system can be described by anomalous viscous hydrodynamics~\cite{Shi:2020htn,Shi:2019wzi} with possible signatures being searched for in heavy ion collision experiments~\cite{Kharzeev:2020jxw}.

So far the discussions focus on the theoretical framework for describing a relativistic rotating fluid, but the primary motivation comes from understanding experimental measurements of spin polarization in heavy ion collisions, currently being carried out at both the Relativistic Heavy Ion Collider (RHIC), the Large Hadron Collider as well as the GSI facility~\cite{STAR:2017ckg,ALICE:2019aid,STAR:2022fan,HADES:2022enx}.
Full development of such a framework will help advance phenomenological studies for interpreting the experimental data.
A recent example is the finding of  non-equilibrium contribution (via thermal shear term) that helps explain puzzling patterns in the local spin polarization data~\cite{Fu:2021pok,Becattini:2021iol}.
Once a consistent and complete framework up to second-order viscous terms is established, it can be applied for simulations of angular momentum transport and computations of various spin polarization observables in heavy ion collisions. A wealth of relevant  data are  anticipated  from ongoing analyses at STAR, ALICE and HADES as well as from planned future experiments such as FAIR.


To conclude, investigation of relativistic fluid with angular momentum emerges as a rapidly advancing frontier with considerable progress and ample opportunities for further developments in theoretical framework, phenomenological modeling as well as experimental measurements. Besides its broad relevance in many-body physics, studies of such a system in nuclear collisions will provide unique insights into one of the most fundamental forces, i.e. the strong interaction as described by Quantum Chromodynamics (QCD). The presence of angular momentum offers a novel probe to help reveal   QCD interactions in the nonperturbative regime. As a famous example, understanding the $\frac{\hbar}{2}$ spin decomposition of a proton (--- a basic building block of the whole atomic world)  in terms of its quark and gluon contents has intrigued physicists for nearly four decades and continues to be a very active area of QCD research. In the case of a proton, the ratio between angular momentum $J$ and baryon number $B$ (both being exactly conserved quantities) is just $J/B=1/2$, while the strong interaction matter created in a heavy ion collision has the same ratio to be several orders of magnitude larger, $J/B\sim 10^{2\sim 3}$. One cannot help but expect fascinating phenomena and important lessons to be found from the heavy ion system and help achieving a much deeper understanding of the ``spinning QCD" physics.

\section*{Acknowledgments}
We thank F. Becattini, X.Y. Guo,  S.Z. Shi, G. Torrieri, H. Yee, H.C Ren for useful discussions and communications. This work is supported in part by the National Natural Science Foundation of China (NSFC) under Grant Nos. 11735007, 11890711, 11890710,12275104， 12205309 by the China Scholarship Council (CSC) Contract No.201906770027, by the Fundamental Research Funds for the Central Universities, as well as by the NSF Grant No. PHY-2209183 and the U.S. Department of Energy, Office of Science, Office of Nuclear Physics, within the framework of the Beam Energy Scan Theory (BEST) Topical Collaboration.

\bibliography{rotation01}

\end{document}